\documentclass{elsart}

\usepackage{psfig}

\usepackage{natbib}

\begin{document}

\runauthor{Oshlack et al.}

% -------------------------------------------------------------------------

\begin{frontmatter}

\title{Unravelling Active Galactic Nuclei}

\author[melb]{A. Y. K. N. Oshlack}, \author[melb]{R. L. Webster},
\author[melb]{M.J. Drinkwater}, \author[melb]{M.T.Whiting},
\author[RSAA]{P.J. Francis}, 
\author[ipac]{F.J. Masci}
\address[melb]{Astrophysics Group, School of Physics, 
University of Melbourne, Parkville 3010, Australia}
\address[RSAA]{MSSSO, Private Bag, Weston Creek, ACT 2611, Australia}
\address[ipac]{IPAC/Caltech, MailStop 100-22, Pasadena, CA 91125, USA}
\begin{abstract}
A complete flat-spectrum radio-loud sample of AGN includes
a significant fraction of Seyfert-like AGN including a NLS1.
Analysis of their optical spectra suggests that the reddest
continuum colours are either associated with AGN in nearby resolved
galaxies, or distant quasars showing relatively narrow
permitted emission lines.
 
\end{abstract}


\end{frontmatter}

% -------------------------------------------------------------------------

\section{Introduction}
The relationship between Seyfert galaxies and radio-loud galaxies has
not been explored in detail. However, a complete sample of flat-spectrum radio-loud AGN
contains a significant fraction of Seyfert-like sources, including one
which would be classified as a narrow-line Seyfert 1 (NLS1).

To understand the physical mechanisms responsible for producing the
different characteristics of AGN, a multiwavelength approach is
needed. Correlations between properties at different wavelengths can be
used to reject and refine physical models of the central regions of
AGN. The Parkes
Half-Jansky Flat-Spectrum Sample (PHFS) \cite{D1} 
is interesting in this
context because it is a radio-selected sample. The selection criteria
are as follows:

\begin{itemize}

\item Radio-loud: 2.7GHz flux $>$ 0.5 Jy.

\item Flat-spectrum: $\alpha_{2.7/5.0} > -0.5$, where
$S_{\nu}=\nu^{\alpha}$

\item Galactic latitude: $|$b$| >20^{\circ}$

\item $-45^{\circ} <$ Dec(B1950) $< +10^{\circ}$

\end{itemize}

This sample contains 323 sources with a wide range of properties which
can be quite different to AGN selected by
optical colours. The PHFS AGN have a large range in optical luminosities, with more than 50 objects having absolute magnitudes
in the Seyfert luminosity range (M$_{B}>$-23). Another
interesting characteristic is the large dispersion
in optical colours of this sample compared with optically selected
samples such as the Large Bright Quasar Survey \cite{W1,F1}. 

\section{Spectra of the Parkes Quasars}

The mechanism responsible for producing
the large dispersion in optical colours, especially the significant
number of red sources in the sample, is not at all well
established. There have been suggestions of dust reddening \cite{W1},
reddening due to the underlying galaxy spectrum \cite{M1} and
synchrotron reddening \cite{W2}. We consider two questions: 
\begin{enumerate}
\item Are the reddened continuum colours associated with changes in the
emission lines of the AGN? 
\item Can we identify the reddening mechanisms
contributing to each AGN? 
\end{enumerate}
We have low resolution optical
spectra of an unbiased subsample of the PHFS. This subsample was
divided into three equal colour bins based on the B-K colours of the
objects: blue (B-K $<$ 3),
intermediate (3 $<$ B-K $<$ 4.4) and red (B-K $>$ 4.4). 
A composite spectrum of the objects in each
bin was made (see \cite{F3} for a description
of the technique used). The results are plotted in Figure
1. It should be noted that the continuum has been normalised to the
same slope for each bin to compare the emission line properties
over the range of colours. The LBQS composite spectrum is also shown
as a comparison with an optically selected sample \cite{F2}.

There are clear differences between the different
composites. The width of H$\beta$ line decreases as redness
increases and O{\footnotesize{II}}(3727\AA) is only seen in the red
composite. The Mg{\footnotesize{II}}(2800\AA) line is also quite broad in the blue and intermediate composites, and much narrower in the red composite.
Further analysis of the spectra in the red bin shows there are two
different types of spectra in this bin: (1) low redshift galaxy-type 
spectra, and (2)
higher redshift quasar spectra. 

Several very red objects in
the Parkes sample seem to appear red owing to a large contribution from
a galactic component. The 4000\AA~break in the spectrum of the galactic
light reduces the amount of
light in the blue filter giving a B-K colour which is quite red. These objects are at low redshifts and
are generally resolved. Masci \cite{M1} has fitted
an evolved elliptical galaxy SED to the PHFS spectra,
showing that these objects have a high proportion of galactic
light in their spectra. These objects also have the
steepest radio spectrum with values of $\alpha  \sim -0.5$. 
The composite spectrum of these resolved sources is fairly
typical of a galaxy with no H$\beta$ emission.

The second group of red objects in the
sample are higher redshift quasar/Seyfert-type objects. They show the strong
emission lines characteristic of quasars and have flatter
radio spectral indices. Whiting et al. \cite{W2} have modelled the reddening 
by synchrotron emission using
broad band SEDs.  They show that these objects tend to
have a strong component of synchrotron emission which is responsible
for their red continuum. 
These quasars would be missed in a
traditional blue-selected quasar sample.
%A composite of the resolved sources in the red subsample shows
%strong O{\footnotesize{II}}(3727\AA) emission, the 4000\AA break,
%and no H$\beta$ emission.  Thus the spectrum is that of a fairly
%typical galaxy.  
The composite of these unresolved sources has a
completely different shape, with narrow permitted emission lines at
CIV(1549\AA), CIII](1909\AA) and MgII(2798\AA) superimposed
on a fairly smooth continuum.  Thus it appears that the red AGN
have relatively narrow emission lines. A fuller description
of these results will appear in Oshlack et al. \cite{O2}.

\begin{figure}[htb]
\centerline{\psfig{file=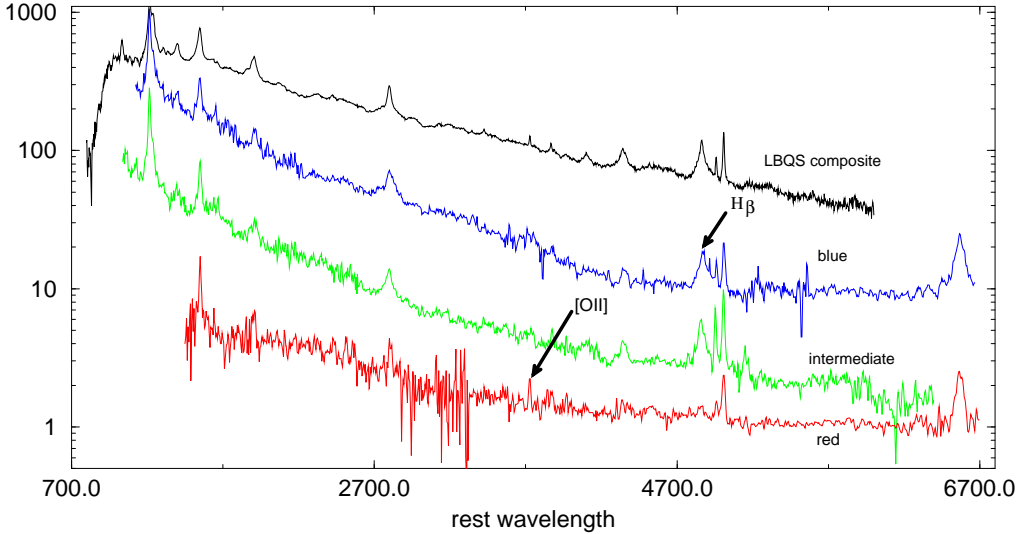,height=3.0truein,angle=-90.0}}
\caption{Composite optical spectra generated from the PHFS
as a function of optical (B-K) colour. The LBQS composite is also
shown for an optically selected comparison. Each composite is normalized to the same continuum slope to compare emission
lines. The plot shows the rest wavelength in angstroms, and the flux in
arbitrary units.}
\end{figure}

\section{A NLS1 in the Parkes Sample}

In examining the spectra from the Parkes sample we have identified a
NLS1 candidate, PKS 2004-447 (Figure 2). One  model for NLS1s suggests they are
AGN oriented towards us, so that the doppler broadening of a disk-like
Broad Line Region would naturally produce narrower emission lines
\cite{O1}. An extremely radio-loud NLS1 
could provide an independent test of such physical models of 
NLS1s, and AGN unification in general. 

\begin{figure}[htb]
\centerline{\psfig{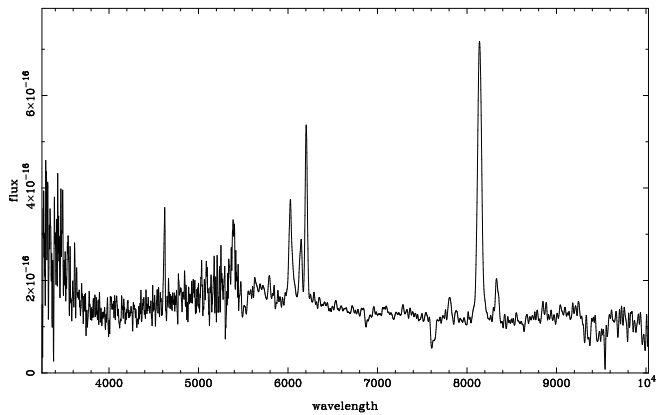}}
\caption{Low resolution spectrum of PKS 2004-447, a candidate radio-loud
NLS1.}
\end{figure}

A low resolution spectrum of this source shows an H$\beta$ width of
$\sim$2000km~s$^{-1}$. There is some indication of Fe{\footnotesize{II}}
emission on the blueward side of H$\beta$ and the source has been detected 
at X-ray energies \cite{S1}. This object is extremely radio-loud with a radio
to optical flux ratio (R) exceeding 7000. The colour of PKS 2004-447 is very
red, in contrast with the previously discovered radio-loud NLS1, RGB J0044+193
which was quite blue \cite{S2}. 
The radio spectrum of PKS 2004-447 is also quite steep, with recent
measurements of the (contemporaneous) radio spectral
index giving $\alpha_{R}=-0.67$, consistent with radio indices of radio-quiet
NLS1s (Moran, these proceedings). A higher resolution spectrum of
PKS 2004-447 will be obtained, to confirm its identity.

% -------------------------------------------------------------------------

% -------------------------------------------------------------------------

\end{document}